\documentclass[9pt,onecolumn,twoside]{opticajnl}
\journal{opticajournal} % use for journal or Optica Open submissions

\usepackage{lineno}
\title{Spectral and spatial filtering of whispering gallery modes in precision-engineered microbubble resonators}

\author[1,2,*]{Ramgopal Madugani}
\author[1]{Amal Jose}
\author[1]{Christophe Pin}
\author[1]{Metin Ozer}
\author[3]{Martina Hentschel}
\author[1,**]{S\'ile {Nic Chormaic}}

\affil[1]{Okinawa Institute of Science and Technology Graduate University, Onna, Okinawa 904-0495, Japan}
\affil[2]{Department of Applied Physics, Waseda University, 3-4-1 Okubo, Shinjuku-ku, Tokyo 169-8555, Japan}
\affil[3]{Institute of Physics, %Faculty of Natural Sciences, 
Technische Universität Chemnitz, Reichenhainer Straße 70, 09126 Chemnitz, Germany}

\affil[*]{ramgopal.madugani@oist.jp}
\affil[**]{sile.nicchormaic@oist.jp}

\begin{abstract}
Similar to microspheres, thin-walled microbubble resonators support whispering gallery modes (WGMs) that combine ultrahigh Q-factors and  small effective mode volumes.
In contrast, their hollow nature enables enhanced interactions with encapsulated materials and lower spectral mode density due to the tight radial confinement of the optical modes. However, the existence of a high axial-mode density still leads to significant mode mixing and modal interference that can complicate spectral shift measurements, thereby limiting sensing applications. To address this limitation, we have fabricated geometric filters directly on the surface of microbubbles using focused ion beam (FIB) milling. Based on  numerical calculations, we first designed and then fabricated large tapered patterns, such as circular surface dips or holes, that could effectively filter modes while minimizing optical scattering losses. Local lateral mode confinement and partial recovery of high Q-factors were experimentally achieved by adding shallow slit patterns. Using few-mode engineered microbubble resonators, we subsequently demonstrated pressure sensing and wide spectral tuning of WGMs free from mode-mixing artifacts. This precision engineering approach promises improved mode isolation, tunable directional emission, and ultrasensitive measurements in microbubble resonator devices.
\end{abstract}

\setboolean{displaycopyright}{false} % Do not include copyright or licensing information in submission.

\begin{document}

\maketitle

\section{INTRODUCTION}\label{sec1}
Optical whispering gallery mode (WGM) resonators have been a research focus for a wide range of applications for decades, especially in photonics, cavity QED, and sensing, due to their ultrahigh Q-factors, small mode volumes, and on-chip integration possibilities \cite{braginsky1989quality,  spillane2003ideality, wu2010ultralow, yan2011packaged, aspelmeyer2014cavity, Fang:17, Kim2019, yu2021whispering, Li2021, zhou2023coupling, xiang20233d, Li2024}. WGM microbubble resonators  \cite{sumetsky2010optical,watkins2011single}, being hollow, offer several advantages over other device types, such as solid microspheres or microdisks. They are also relatively easy to fabricate, can be engineered for mode dispersion  and nonlinear effects \cite{yang2016degenerate, Yang2016Four-wave, Riesen2016Dispersion, Tian2022},  can  contain plasma for microphotonics applications \cite{bathish2023absorption}, are pressure-tunable \cite{yang2015coupled-mode-induced,madugani2016linear,yang2016high-q}, can be used for \textit{in-situ} particle sensing and spectrometry \cite{Ward2018Nanoparticle,hogan2019toward}, and as force sensors \cite{li2024all,shen2025enhanced}. Due to their hollow nature, microbubbles are especially useful for sensing applications, as they can confine fluids and particle solutions  \cite{yang2014quasi,Ward2018Nanoparticle,pan2022active}; this can facilitate sensing and/or particle trapping with minimal environmental disturbance, as demonstrated for magnetic field and magnetobacterial sensing  \cite{liu2021magnetic, Amal2024Magnetospirillum}. Furthermore, similar to 3D chaotic microspheres \cite{tian2025x}, microbubbles may offer a unique opportunity to create precision hollow chaotic microcavities and additional shape modifications may be exploited to observe optical spin-orbit coupling \cite{ma2016spin,martina2020spinorbit}. 

Despite these possibilities, owing to their bottle-like profile and rotational asymmetry, microbubbles possess a dense spectrum of WGMs with similar radial and azimuthal, but different polar, distributions of the light field. Axial mode confinement and spectral control have been extensively explored in several platforms such as surface nanoscale axial photonics (SNAP) devices, where nanoscale effective radius variations enable deterministic axial mode engineering \cite{Sumetsky:11}. However, translating such axial-mode control methods to hollow resonators is challenging due to their curved thin-walled geometry and sensitivity to surface perturbations. Empty microbubbles may already provide some filtering of higher-order radial modes, if the wall is so thin that it acts as a planar slab waveguide \cite{yang2014quasi,Jalaludeen2023Structural,korenjak2024smectic}. For wall thicknesses close to the light wavelength, light confinement between both the inner and outer surfaces stipulates that only a single radial mode can survive if the refractive index of the internal environment is air (or vacuum). Thinner walls  eventually lead to significantly degraded quality factors, as modes can no longer be efficiently confined in the silica layer \cite{yang2014quasi}. Higher-order, high-Q radial modes can be recovered by filling the cavity with fluids of refractive index  between those of air and silica, such as water, thereby entering the so-called quasi-droplet regime \cite{yang2014quasi,Ward2018Nanoparticle}. 

The spectral response of typical microbubble resonators exhibits numerous occurrences of mode mixing and interference, resulting in complex spectral patterns that hinder reliable single-mode tracking for sensing applications. This has been an obstacle for some of the aforementioned applications, where a single high-Q mode isolated in a broad spectral window would be preferred.  Recently, Rodemund et al. introduced the Husimi functions for three-dimensional cavities to provide a deeper theoretical understanding of the mode morphology and  dynamics, including polarization dependence \cite{Rodemund25}. Experimental progress on mode control in microsphere resonators is significantly more advanced.   For example, filtering or selecting specific modes of microspheres  was attempted by depositing an index matching liquid at an off-equator location \cite{SenthilMurugan2011Hollow-bottle} and by engraving the surface of a resonator using a focused ion beam (FIB) \cite{ding2012whispering} or a femtosecond laser \cite{xiang2024spectral}. Off-centered patterns drilled on the side of the light path introduced a perturbation that selectively degraded the resonance of some higher-order axial modes. By doing so, one could efficiently filter out those unwanted modes. However, the isolated modes also suffered from high losses and Q-factor degradation after surface patterning. This limitation can be addressed by redirecting the light path of the selected modes to minimize their interaction with the filtering pattern. To this end, transformation optics and metamaterial-based approaches are especially promising. For instance, Lee et al. demonstrated that spatially varying the nanohole density distribution in InGaAsP microdisk lasers successfully recovered  the Q-factor following its degradation by boundary shape deformations \cite{Lee:23}.  

Precision device engineering is, therefore, of fundamental interest for both improving the performance of microbubble resonators and facilitating their implementation in practical devices. As mentioned, post-fabrication spectral control using FIB processing has been demonstrated in solid microresonators, enabling spectral trimming and resonance alignment through controlled surface modification \cite{ding2012whispering, Yin:18}. These approaches have primarily targeted azimuthal or radial mode tuning or frequency correction in solid devices, rather than deterministic axial-mode selectivity in hollow resonators. In this paper, we present precision fabrication methods using FIB milling to achieve selective axial WGM isolation and optimization, both spatially and spectrally, in hollow cavities. We further demonstrate the modified microbubble's sensing capability by using the filtered devices for pressure sensing and tuning, where the WGMs show no mode mixing or distortions throughout. Using this precision fabrication method to pattern the microbubbles, we experimentally demonstrate the realization of tightly localized intensity spots and partial Q-factor recovery.

\section{NUMERICAL ANALYSIS}\label{sec2}

Microbubbles with diameters of several tens of wavelengths allow light to be guided in silica shells with wavelength-scale wall thicknesses, enabling tight radial confinement and small mode volumes while preserving high Q-factors. Unfortunately, performing precise 3D numerical simulations of their optical response is often too resource intensive. Hence, we have restricted our study to axisymmetric resonators, allowing  us to only consider their transverse cross-sections. Using COMSOL Multiphysics, we calculated and compared the WGMs of microbubble resonators milled 15~{\textmu}m away from the equator on one side with different milling depths from zero to complete milling. The geometrical parameters were chosen according to measurements performed on one of the experimentally fabricated and characterized microbubble resonators. In the equatorial plane, the microbubble was assumed to have a diameter of 155~{\textmu}m and a wall thickness of 0.84~{\textmu}m, with a transverse curvature radius of 190~{\textmu}m. The increase in wall thickness away from the equator was calculated following the conserved silica volume approximation \cite{Jalaludeen2023Structural}.

\begin{figure*}[htb]
\centering\includegraphics[width=0.9\linewidth]{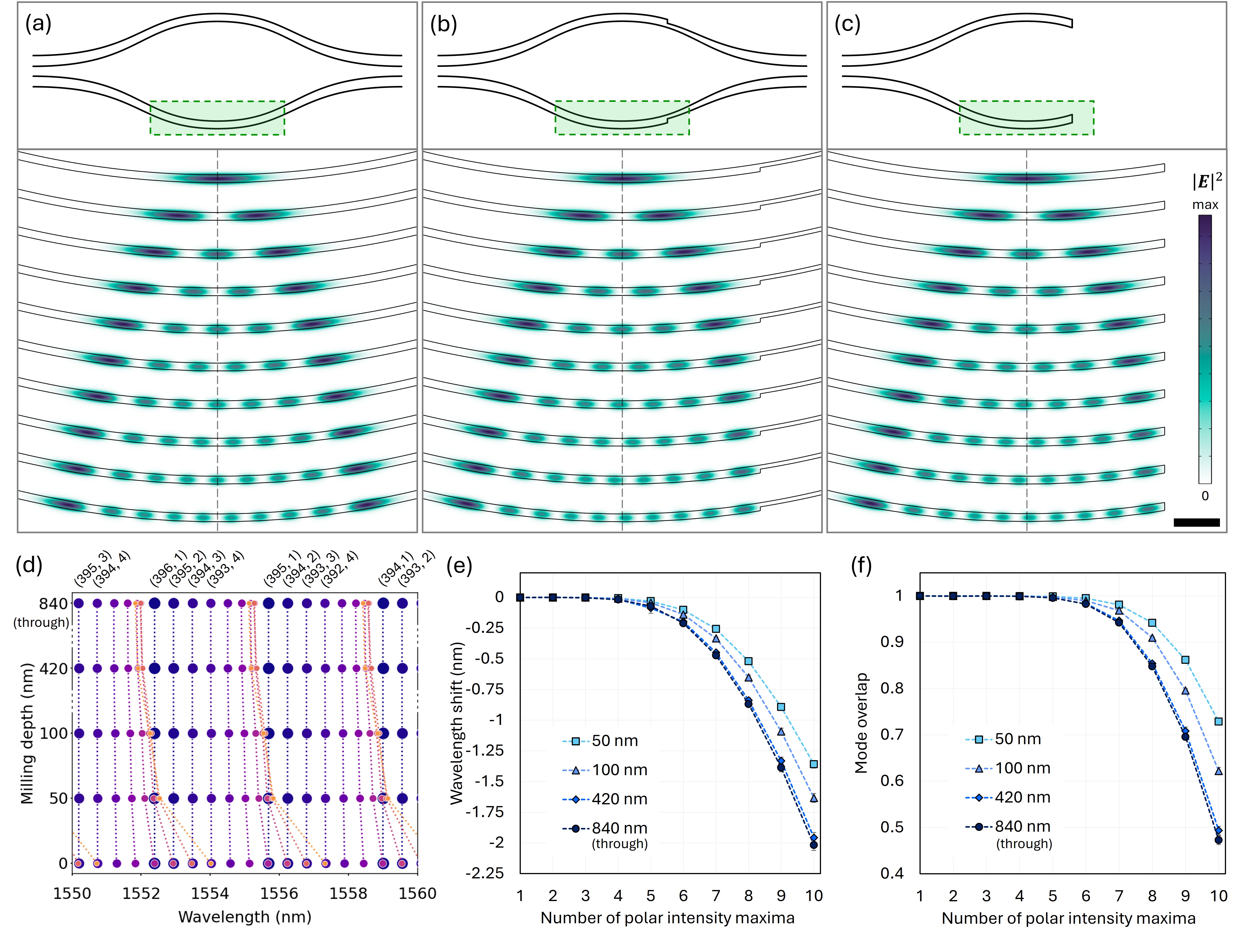}
\caption{Simulated intensity distribution of the first ten TM WGMs with an azimuthal mode number of 293 for a 155~{\textmu}m diameter microbubble (a) without modification, (b) partially milled, and (c) completely truncated on one side (as depicted in the cross-section schematics). Scale bar: 5~{\textmu}m. (d) Resonant wavelengths of the WGMs with up to ten intensity maxima along the polar direction in the 1550~nm to 1560~nm range. The data points are sorted by colors and sizes (descending order) according to the number, $p$, of polar intensity maxima (i.e., the axial order). For the lowest-order axial WGM, the azimuthal mode numbers, $m$, and $p$ are reported above the corresponding curves. (e) Resonant wavelength shifts and (f) normalized mode overlap integral values as a function of the milling depth (averaged values for azimuthal mode numbers from $m = 293$ to $m =297$).}
\label{Figure1ModeFilterCOMSOLAnalysis}
\end{figure*}

Figures \ref{Figure1ModeFilterCOMSOLAnalysis}(a)-(c) illustrate the case of an unmilled microbubble, a partially milled microbubble and a completely milled (effectively truncated) microbubble, respectively. The resonator only supports confined WGMs with a radial number, $n = 1$. Multiple WGMs may share the same polar quantum mode number, $l$, but have different azimuthal quantum mode numbers, $m$ (and vice versa). $l$ and $m$ relate to the quantized total angular momentum carried by the WGM and its quantized projection onto the symmetry axis of the cavity, respectively. While $m$ directly gives the number of periods in the WGM field along the azimuthal direction, the number of intensity maxima, $p$, along the polar or axial direction is equal to $p = l - |m| +1$ \cite{yu2021whispering}. In this work, $p$ is used to define the polar order of a WGM. The intensity profiles of the first ten transverse-magnetic (TM) WGMs with the same azimuthal mode number, $m =$ 293, are shown. From the sixth-order axial mode, even a shallow carving results in a clear asymmetry in the intensity distribution, see Fig. \ref{Figure1ModeFilterCOMSOLAnalysis}(b). This spatial energy redistribution occurs in conjunction with a spectral shift of the resonant wavelength. Figure \ref{Figure1ModeFilterCOMSOLAnalysis}(d) shows the resonant wavelengths of the WGMs with up to ten polar intensity maxima along the direction of the symmetry axis that lie between 1550~nm and 1560~nm. A clear wavelength shift appears for WGMs with more than five polar intensity maxima. Figures \ref{Figure1ModeFilterCOMSOLAnalysis}(e) and (f), respectively, summarize the wavelength shifts  and the normalized mode overlap integral  values \cite{wiersig2011nonorthogonal} averaged over the WGMs with azimuthal mode numbers ranging from $m = 293$ to $297$. The normalized overlap integrals were computed as follows:
\begin{equation}
    S = \frac{\big|\iint_{A} \psi_{0}^* \psi \,dr\,dz \big|}{\sqrt{\iint_{A} \psi_{0}^* \psi_{0}^{ } \,dr\,dz}\sqrt{\iint_{A} \psi^* \psi \,dr\,dz}},
\end{equation}
where $\psi_0$ is the reference mode field distribution (i.e., the WGMs of the original microbubble), $\psi$ is the field distribution of the WGM under consideration, and $A$ is a sufficiently large cross-sectional area of the resonator. Despite the lack of rotational symmetry of patterned microbubble resonators, these results provide a useful estimation of the mode order-dependent strength of the perturbation caused by a large hole drilled on the surface of a resonator. Significant degradation of the Q-factor or even complete filtering out is expected for WGMs with more than four  axial intensity maxima. For a more quantitative analysis of the filtering effect, the scattering efficiency of the hole pattern interacting with the guided light should be evaluated by considering radiation loss channels both inside and outside the silica microbubble.

 \section{EXPERIMENTAL METHODS AND RESULTS}\label{sec3}
 
Spatial filtering and confinement of WGMs was experimentally investigated by using FIB-processed microbubble cavities. The microbubble cavities were prefabricated using a focused C$\text{O}_\text{2}$ laser on tapered silica glass capillaries while pressurizing with $\text{N}_\text{2}$ gas {\cite{sumetsky2010optical,watkins2011single,ward2013highly,madugani2016linear}. Large off-equator rectangular or circular patterns were milled into the microbubbles to achieve significant higher-order axial mode filtering. Furthermore, shallow slit patterns were milled around the equator to achieve local spatial mode confinement. Spectral and spatial characterizations of the WGMs were performed by coupling laser light to the cavity through a fiber taper, analyzing the fiber-transmitted signal, and imaging the radiated light. For further details, see the Appendix.  

\subsection{Mode Filtering by Rectangular Surface Milling}\label{subsec3.1}

\begin{figure*}[htb]
\centering{\includegraphics[width=0.78\linewidth]{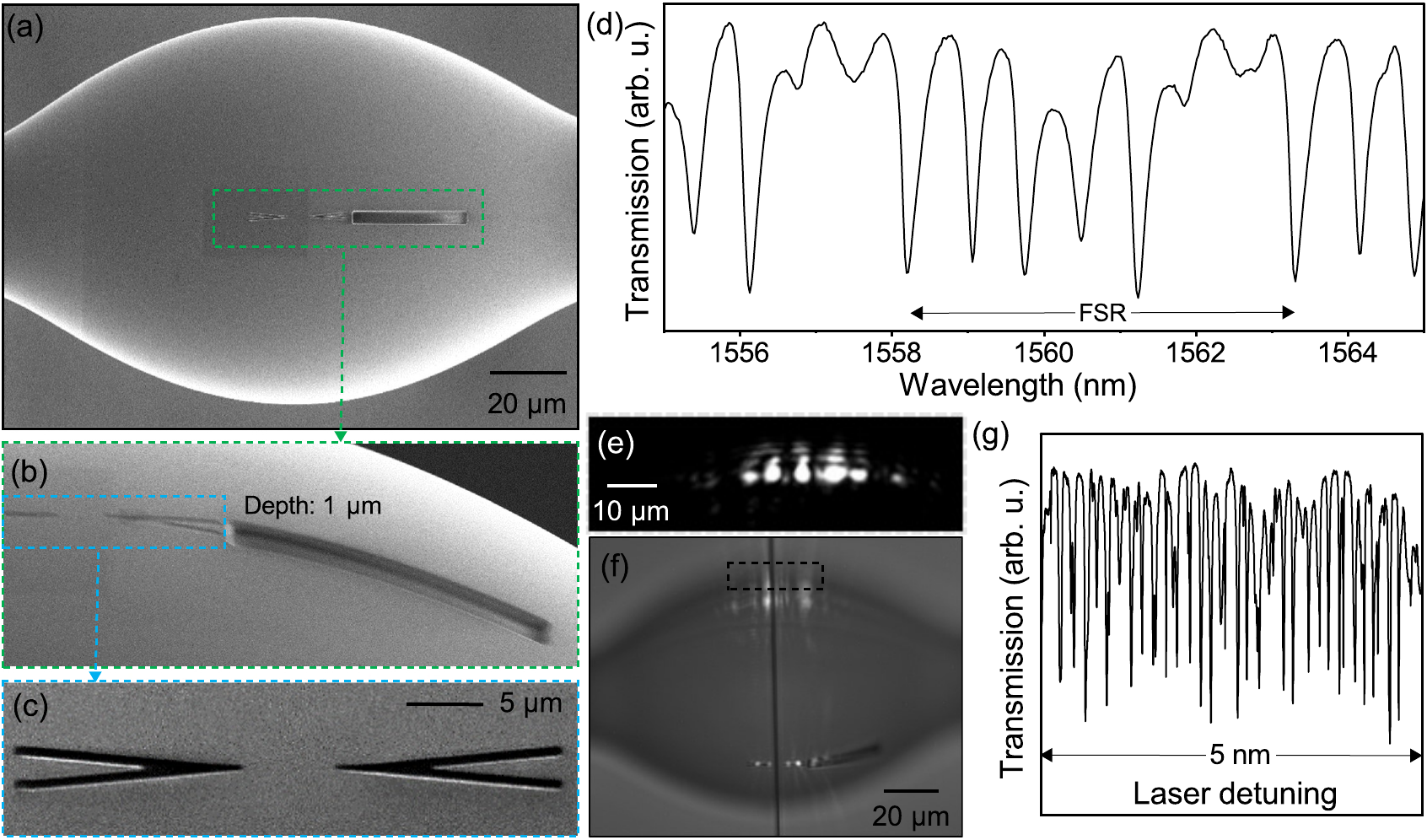}}
\caption{WGM filtering effect of a large rectangular dip. (a) Scanning electron microscope (SEM) image of a 101~{\textmu}m diameter microbubble, with a wall thickness of 1.5~{\textmu}m, after  milling an off-equator $5 \times 30${~{\textmu}m}$^2$ rectangle next to a 400 nm deep "><"-like slit pattern spanning the equator. The dotted  area in (a) is partially shown from an oblique angle in (b), revealing a milled depth of about 1~{\textmu}m for the rectangular dip. The dotted  area in (b) is shown in (c) as a zoom of the "><"-like slit pattern composed of four 500 nm-thick groove lines,  tapered down to about 100 nm as they meet, with a 5~{\textmu}m gap, at the equator. (d) Filtered WGM spectrum in a two-FSR window, with  Q-factors not exceeding $1\times10^4$. Field distribution of the fifth-order axial mode excited at 1556.13 nm via a tapered fiber as revealed by IR camera images of the light radiated (e) from the top contour and (f) from the milled area of the microbubble. (g) WGM spectrum of the unmilled microbubble over one FSR  with a measured Q-factor of  $2\times10^5$.}
\label{Figure2RectangularSurfaceMillandSpectrum}
\end{figure*}

A large rectangular dip, with dimensions $5 \times 30$ {~{\textmu}m}$^2$, in combination with a shallow "><"-like slit pattern were milled on the surface of a 101{~{\textmu}m} diameter microbubble. As shown in Figs. \ref{Figure2RectangularSurfaceMillandSpectrum}(a,b), the  rectangle was milled 15~{\textmu}m away from the equator. Considering an initial wall thickness close to 1.5~{\textmu}m, a milling depth of 1~{\textmu}m  was chosen to  prevent local light propagation in the remaining 500-nm-thick milled area \cite{yang2014quasi,Ward2018Nanoparticle}. A milling depth of 400~nm  was chosen for  the slit pattern composed of 500~nm-wide grooves and a 5~{\textmu}m-wide gap between both slits, see Fig. \ref{Figure2RectangularSurfaceMillandSpectrum}(c). The milling of the  rectangular dip  resulted in a significant reduction in the number of WGMs present. Figure \ref{Figure2RectangularSurfaceMillandSpectrum}(d) shows that only modes up to the fifth axial order  were excited post-milling. The broadening of the resonance dips illustrates that the Q-factor also dropped by about 25$\times$ that before milling, see Fig. \ref{Figure2RectangularSurfaceMillandSpectrum}(g). The initial microbubble spectral response shows a dense mode distribution over a typical free-spectral-range (FSR) window of 5.05 nm. Localized geometric perturbations are known to induce avoided crossings, mode hybridization, and selective enhancement or suppression of cavity eigenmodes in deformed WGM systems \cite{Wiersig.97.253901}. In our case, the FIB-induced dip acts as a spatially localized perturbation that selectively modifies the axial field distribution while preserving the primary azimuthal confinement. Figures \ref{Figure2RectangularSurfaceMillandSpectrum}(e) and (f) show infrared (IR) images of the light radiated by the fifth-order axial mode (Q-factor of 7600) excited at 1556.13 nm wavelength. The mode intensity distribution extends for approximately 15~{\textmu}m  either side of the equator. As predicted by our simulations, higher-order modes with more extended intensity profiles are  filtered out because of their increased interaction with the large rectangle. We also verified that the slit pattern alone does not lead to any significant mode filtering or broadening of the WGM resonances. The resonant mode degradation can be explained by the increase in scattering losses caused by the deeper, and several  wavelengths larger, rectangular dip that acts as a knife edge, partially cutting one side of the guided light beam.

\subsection{Mode Filtering by Circular Surface Milling}\label{subsec3.2}

\begin{figure*}[htb]
\centering\includegraphics[width=0.78\linewidth]{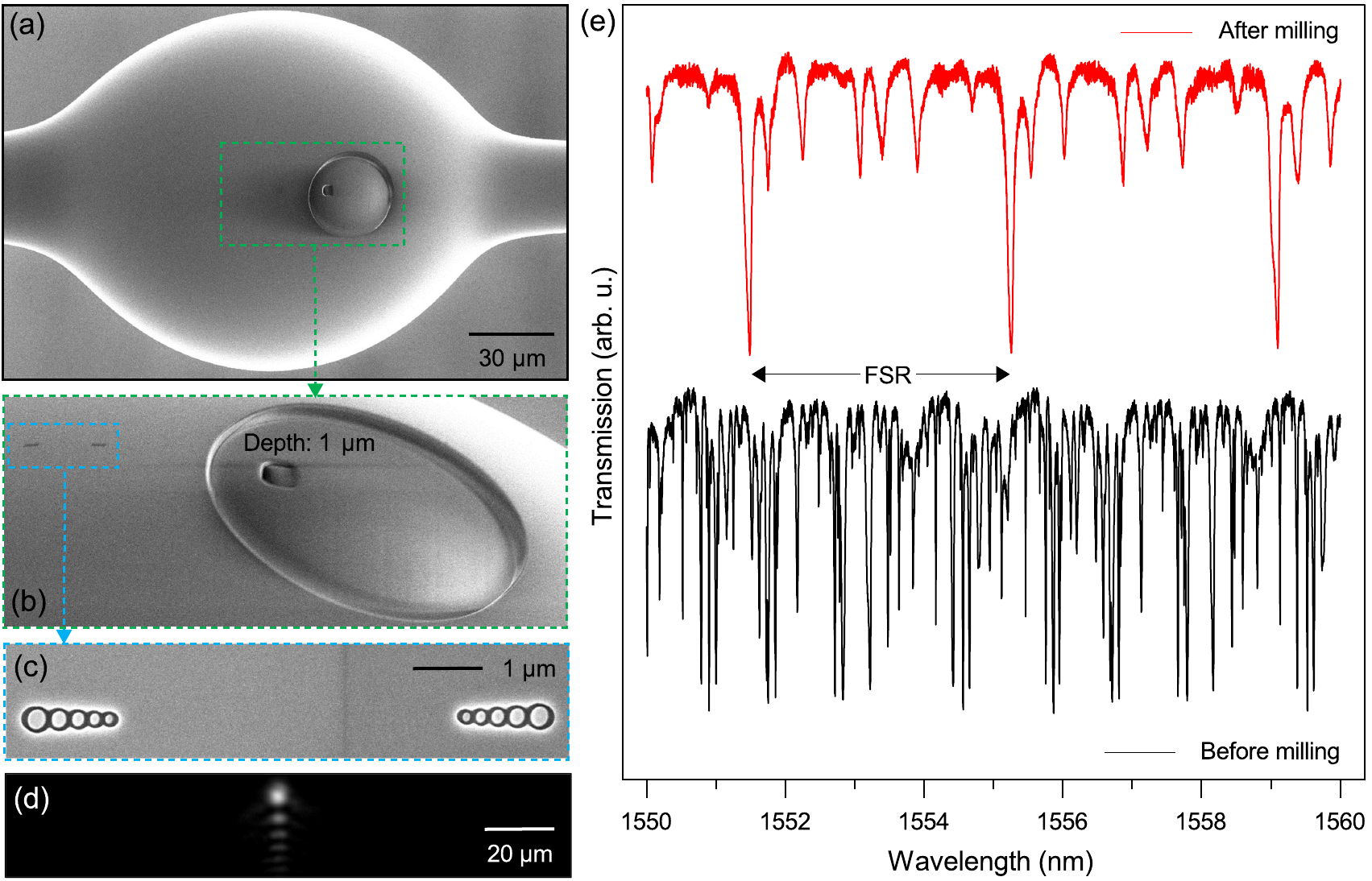}
\caption{WGM filtering effect of a large circular dip. (a) SEM image of a 130~{\textmu}m diameter microbubble, with a wall thickness of 1.3~{\textmu}m, after milling an off-equator 30~{\textmu}m diameter disk next to a shallow slit pattern at the equator. The 100~nm-deep, 4~{\textmu}m$^2$ surface mill within the disk was used for finer FIB focus adjustment. The dotted box in (a) is shown from an oblique angle in (b), revealing a milled depth of $\sim$1~{\textmu}m for the disk. The dotted box in (b) is shown in (c) as a zoom on the shallow slit pattern composed of 300 nm-deep and no more than 100 nm in groove width annuli of increasing diameters (an average of inner and outer part of groove) varying from 100 nm to 500 nm, with a 5~{\textmu}m gap at the center. (d) Field distribution of one of the filtered fundamental axial modes excited at 1551.4 nm via tapered fiber coupling. The IR camera focus was set near the shallow milled patterns. (e) Filtered WGM spectrum (red) over a few-FSR-wide spectral window (3.87 nm FSR) in comparison with the  spectrum of the unmilled microbubble (black). The fundamental mode has a Q-factor of $1.6\times10^4$, one order of magnitude below the Q-factor of the unmilled microbubble of $1.3\times10^5$.}
\label{Figure3CircularSurfaceMill}
\end{figure*}

To reduce  scattering losses caused by the sharp edges of the rectangular dip, the interaction with the lowest-order axial modes can be reduced by tapering the pattern close to the light path, i.e. near the equator. To demonstrate this point, a 30~{\textmu}m-diameter and 1~{\textmu}m-deep circular dip was milled 15~{\textmu}m from the equator on a 130~{\textmu}m diameter resonator with wall thickness $\sim$1.3~{\textmu}m, see Figs. \ref{Figure3CircularSurfaceMill}(a) and (b). As shown in Fig. \ref{Figure3CircularSurfaceMill}(c), a shallow slit pattern consisting of 300 nm-deep adjacent annuli of increasing diameters ranging from 100 nm to 500 nm in 100 nm increments with a 5~{\textmu}m-wide gap was also milled to help preserve the Q-factor of the lowest-order modes. 

After filtering, a fundamental TE mode at 1551.4 nm, with a confined intensity field distribution, was imaged using an IR camera focused near the shallow pattern area, as shown in Fig. \ref{Figure3CircularSurfaceMill}(d). Figure \ref{Figure3CircularSurfaceMill}(e) shows the corresponding spectral responses of the microbubble before and after FIB milling. For details, see the Appendix. Significant mode filtering was achieved, leaving mostly only the fundamental mode with a Q-factor of $1.6\times10^4$ and an FSR of 3.87~nm. Despite mode degradation, the drop in Q-factor was less than an order of magnitude, e.g., 8 times lower for the fundamental mode. Large circular patterns thus appear to be suitable for efficient mode filtering with minimal mode degradation. The influence of the milling depth was investigated using separate microbubble samples of similar dimensions. Both shallow-cut (500 nm-deep) and complete through-cut milling attempts resulted in mode filtering, the former leading to slightly larger drops in Q-factors compared to the latter. Further examples of both cases are presented in the following sections.

\subsection{Optimized Spectral Filtering by Local Mode Confinement}\label{subsec3.3}

As previously discussed, a shallow slit pattern may help preserve the high Q-factors of the lowest-order WGMs, in particular of the fundamental mode, by locally confining the transverse intensity distribution of the modes. A microbubble of 155~{\textmu}m in diameter, $\sim$0.84~{\textmu}m wall thickness was used to demonstrate this effect, see Figs. \ref{Figure4FilterAndConfinement}(a)-(f).   A 30~{\textmu}m-diameter circular hole was fully milled through the microbubble wall at 15~{\textmu}m from the equator to ensure efficient mode filtering, see Figs. \ref{Figure4FilterAndConfinement}(a,b).  We  chose a shallow slit pattern with a combination of adjacent annuli with incremental diameter changes from 100 nm to 500 nm, see Fig. \ref{Figure4FilterAndConfinement}(c), with a slit gap of 5~{\textmu}m across the equator. This design is similar to that shown in Fig. \ref{Figure3CircularSurfaceMill}(c).  To increase the influence of the milled structures, we  extended the slit by including an "><" -like profile, similar to that of Fig. \ref{Figure2RectangularSurfaceMillandSpectrum}(c). To ensure minimal surface modification, the "><" -like profile here follows a framed line milling composed of tilted rectangular frames, 9~{\textmu}m-long and 500~nm-wide, with a groove width of 50 nm and 400 nm deep, milled on both sides of the equator, next to the circular hole. 

As shown in Fig. \ref{Figure4FilterAndConfinement}(d), the initial spectral response of the unmilled resonator revealed a high density of WGMs with Q-factors up to $7.8\times10^5$. This spectrum was obtained when the fiber taper was in contact coupling; hence, the intrinsic Q-factors of the WGMs may be even higher. The spectral response of the cavity was also recorded  after applying a 50 nm-thick ITO coating, see Fig. \ref{Figure4FilterAndConfinement}(e). After the coating, the lower WGM density, while still maintaining a high Q-factor ($1.5\times10^5$), reveals a relatively small mode degradation, with the reduction to Q being less than an order of magnitude.  Post-milling,  efficient and high-Q-preserving mode filtering was achieved, leaving only the resonant dips of the four lowest-order modes in the recorded spectrum, see Fig. \ref{Figure4FilterAndConfinement}(f).  This is in good agreement with the simulation results presented in Fig. \ref{Figure1ModeFilterCOMSOLAnalysis}. With a Q-factor of $8.7\times10^4$, indicating a 1.7$\times$ reduction, preservation of the fundamental mode marked as $l=m$ at 1555.5 nm in Fig. \ref{Figure4FilterAndConfinement}(f), was particularly noticeable due to the influence of the shallow slit pattern. The other higher-order axial modes, indicated by resonant dips  on the right-hand side of this fundamental mode, also maintained relatively high Q-factors ($2.1\times10^4$, $1.5\times10^4$, and $1.6\times10^4$, respectively).

\begin{figure*}[htb]
\centering\includegraphics[width=\linewidth]{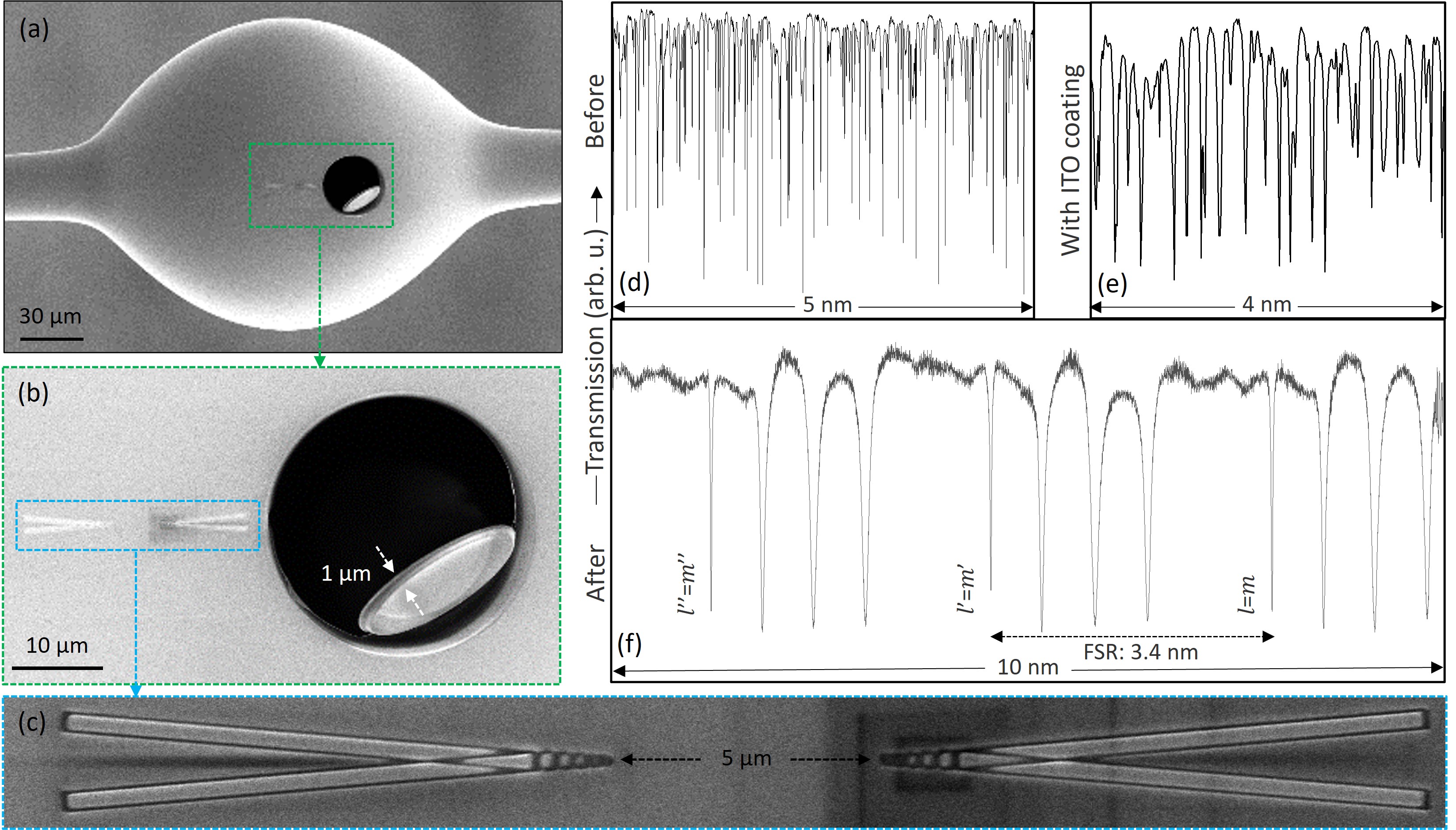}
\caption{High-Q filtered WGMs. (a) SEM image of a 155~{\textmu}m-diameter microbubble resonator, with a wall thickness of 0.84~{\textmu}m, after milling an off-equator 30~{\textmu}m-diameter circular hole. (b) Zoom on the dotted box area in (a) showing the 1~{\textmu}m-thick silica disk partially released during the FIB processing but still attached on one side. (c) Zoom on the dotted box area in (b) showing the shallow 400~nm-deep slit pattern composed of tilted "><"-like rectangular frames terminated with adjacent annuli of increasing diameters with a 5~{\textmu}m gap at the equator. (d,e) Original  WGM spectra (d) before and (e) after ITO coating. (f) Filtered mode spectrum over a 10~nm window around 1550~nm  measured after FIB processing, revealing sharp resonant dips. The measured FSR of 3.4~nm and relative mode spacings (0.633 nm, 0.621 nm, and 0.627 nm) are in good agreement with the simulation results presented in Fig. \ref{Figure1ModeFilterCOMSOLAnalysis}. The fundamental modes are indicated by $l=m, m',... .$}
\label{Figure4FilterAndConfinement}
\end{figure*}

\begin{figure*}[htb]
\centering\includegraphics[width=\linewidth]{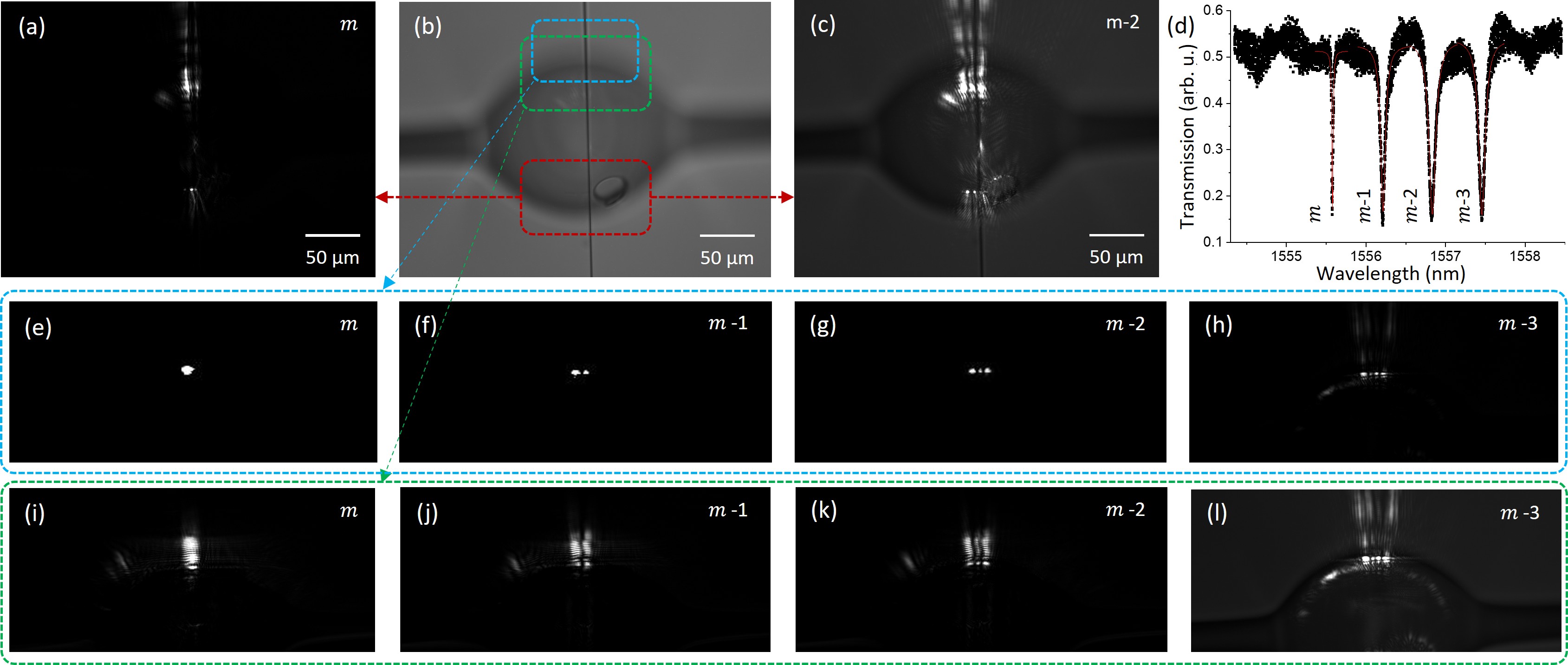}
\caption{Field distributions of the filtered WGMs. The IR camera images show the light radiation scattered at the submicron pattern focus (a) for  the fundamental mode, (b) highlighting different focus positions of the IR camera, (c) for the third-order axial mode, of the spectrum shown in (d). IR camera images for focusing on the top-most edge of the microbubble for (e) the fundamental mode, (f) the second-, (g) the third-, and (h) the fourth-order axial modes with decreasing azimuthal order. (i) - (l) IR camera images of the mode patterns  with intermediate focus between the top and fiber focus point, for the fundamental mode to the fourth-order axial mode, respectively.}
\label{Figure5Localization}
\end{figure*}

Identification of the order of the filtered WGMs  was achieved by imaging the spatial intensity distribution of each mode using an IR camera, see Fig. \ref{Figure5Localization}. The tunable laser was locked to each of the filtered mode's eigenfrequencies, and the light scattered from the cavity to the far-field was imaged. The camera was  focused either on the milled pattern [Figs. \ref{Figure5Localization}(a),(c)], on the top contour of the resonator [Figs. \ref{Figure5Localization}(e)-(h)], or on an intermediate position [Figs. \ref{Figure5Localization}(i)-(l)]. In particular, the intensity distribution of the fundamental mode appears more tightly confined at the shallow slit location, see Fig. \ref{Figure5Localization}(a), than at other locations, see Figs. \ref{Figure5Localization}(e) and (i). In these cases, the limited effect of the shallow slit may be explained by the more extended mode profiles of the higher-order axial modes, leading to a larger overlap with the slit pattern, hence increased scattering losses.

\subsection{Wide Spectral Range Pressure Tuning of Microbubble WGMs}\label{subsec3.4}

To highlight the advantage of precision-engineered microbubble cavities, a partially milled cavity was used to demonstrate pressure-tuning of the filtered modes' eigenfrequencies \cite{madugani2016linear}. A microbubble similar to those already mentioned, i.e.,  156~{\textmu}m in diameter, with estimated wall thickness of 1.1~{\textmu}m, was fabricated and coated with no more than a 50 nm thin layer of ITO. We kept a long section of untapered capillary on one side so that the microbubble could be connected to a pressurized nitrogen gas supply. As shown in Fig. \ref{Figure6PressureTuningwithModeFilters}(a), the WGMs of the ITO-coated cavity were excited. The radiated intensities of the first few orders of axial WGMs were imaged %s %change by MH
by successively locking the laser to the WGMs' eigenfrequencies. The spectral response of the ITO-coated resonator was also recorded prior to FIB milling, revealing Q-factors of $\sim$$10^6$, see Fig. \ref{Figure6PressureTuningwithModeFilters}(b).

Following the  method  described in Section \ref{subsec3.2}, a 30~{\textmu}m-diameter and 500 nm-deep disk was  milled 15~{\textmu}m from the equator to filter out any higher-order axial modes, Fig. \ref{Figure6PressureTuningwithModeFilters}(c). As shown in Fig. \ref{Figure6PressureTuningwithModeFilters}(d), only a few resonances remained in the spectral response of the resonator. Despite increased scattering losses and thermal broadening \cite{carmon2004dynamical}, the fundamental axial modes, i.e., the largest resonant dips in the spectrum, have Q-factors on the order of $10^4$. Partial recovery of the high-Q resonances was achieved by milling an additional shallow slit (400 nm-deep, with adjacent annuli of increasing diameter) around the equator, as shown in Fig. \ref{Figure6PressureTuningwithModeFilters}(e), which also shows the TM and TE field distributions. The TM and TE WGM spectra of the engineered resonator are shown in Figs. \ref{Figure6PressureTuningwithModeFilters}(f) and (g), respectively. Notably, the slit pattern also improved the resonance of the first few axial modes, yielding up to three times higher Q-factors, as revealed by the deeper and sharper dips observed in both spectra. It can be noticed that the milling of the shallow slit also led to a slight resonance shift on the order of 0.1 FSR.

\begin{figure*}[htb]
\centering\includegraphics[width=0.9\linewidth]{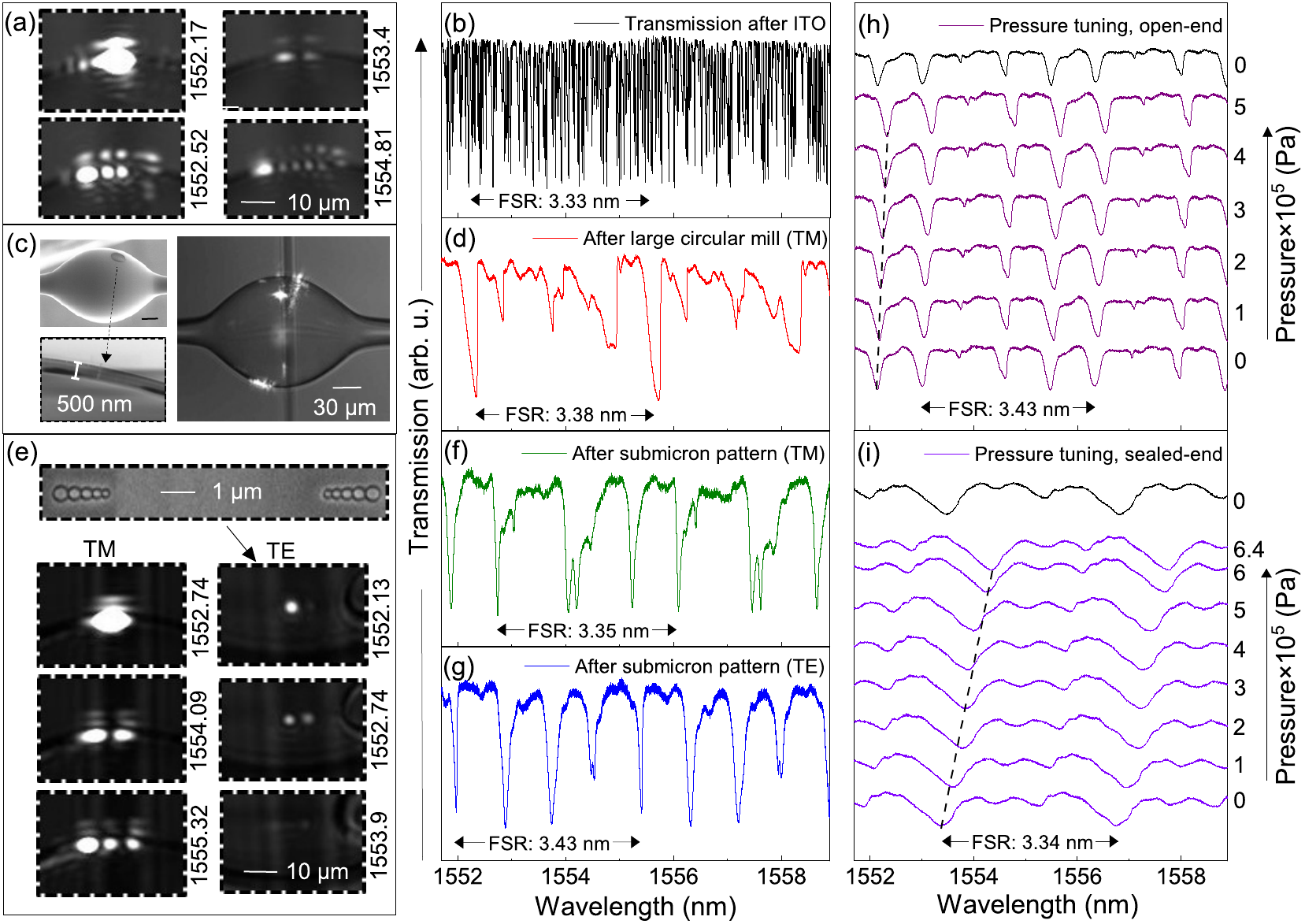}
\caption{Pressure tuning of the WGMs of a partially milled microbubble cavity. (a) IR camera images showing the intensity distribution of the first few axial  modes of an ITO-coated microbubble prior to FIB processing. (b) Corresponding dense WGM spectrum revealing modes with Q-factors of about $10^6$ around 1552 nm. (c) SEM and IR camera images of the microbubble after the fabrication of a 30~{\textmu}m-diameter and 500 nm-deep circular mill, 15~{\textmu}m from the equator. (d) Corresponding filtered WGM spectrum. The fundamental axial mode sustains a Q-factor of approximately $1\times10^4$ with an FSR of 3.38 nm. (e) SEM image of a shallow (400 nm-deep) slit pattern at the equator and IR camera images showing the intensity distribution of the first few axial TM and TE modes. (f)\&(g) Corresponding filtered WGM spectra. The fundamental axial modes appear to have recovered to a slightly higher Q (TM: $2.5\times10^4$; TE: $3\times10^4$) compared to the spectrum in (d). Measured FSRs of the TM and TE modes of 3.35 nm and 3.43 nm, respectively. Recorded spectra in the (h) open-end and (i) sealed-end scenarios of pressure tuning (or sensing) using the engineered microbubble resonator. The dashed lines indicate the fitted slopes revealing pressure sensitivities of 79 GHz/MPa and 154 GHz/MPa, respectively.}  
\label{Figure6PressureTuningwithModeFilters}
\end{figure*}

To demonstrate pressure tuning of the filtered WGM resonances, one of the untapered capillary ends was attached to a N$_2$ gas supply. As shown in Fig.  \ref{Figure6PressureTuningwithModeFilters}(h), the pressure-dependent spectral shift of the WGM resonances was monitored while pressurizing the inner hollow volume of the microbubble from 0 to 0.5 MPa (max relative pressure) in $1\times10^5$ Pa increments. The experiment was repeated after sealing the other end of the capillary using UV-cured epoxy, see Fig.  \ref{Figure6PressureTuningwithModeFilters}(i). In the first case (open capillary end), a linear regression of the spectral shifts measured for a few different axial modes led to an average input pressure sensitivity of 79 GHz/MPa. However, for the second scenario, with a sealed capillary end, a higher pressure sensitivity of 154 GHz/MPa was obtained. A negligible shift rate deviation was observed between modes of different axial orders. Note that, despite the significant mode degradation caused by the sealing process, it was still possible to track the resonant shift of the surviving WGMs.

The demonstrated sensitivities are consistent with our previous work on unprocessed microbubble cavities, but offer a lower limit-of-detection (LOD) due to the lower Q-factors. Furthermore, the pressure-based shifted mode spectrum  shows neither mode mixing nor interference effects over the pressure tuning range (0.1 FSR), hence demonstrating the usefulness of these mode-filtered microbubbles as highly tunable cavities.

\section{DISCUSSION}\label{sec4}

It is widely acknowledged that any surface modification of WGM microcavities leads to degradation of the WGM resonances \cite{foreman2015whispering,jiang2020whispering}. For this reason, surface patterning has often been disregarded as a viable solution for efficient mode filtering in high-Q WGM resonators. However, this work demonstrates that efficient mode filtering can be achieved while minimizing the scattering losses caused by surface patterning. Several strategies may be followed to improve the mode selectivity of the filtering pattern  or to recover, even partially, the Q-factors of the filtered modes. In the first case, tapered pattern designs (such as large circular dips) can be adopted to reduce the impact of the optical filter on the lowest-order WGMs. In the second case, transformation optics-inspired methods can be used to tune the effective refractive index of the resonator's material \cite{Lee:23}. Alternatively, small footprint optical elements (such as shallow diffracting slit patterns) may be fabricated on the surface of the resonator to tune the light path of the WGMs while introducing minimum losses.

Due to their hollow structure, microbubble resonators already sustain WGMs with only a few, if not a single, radial orders. According to the experimental results reported here, efficient filtering of the high-order axial modes can also be achieved by milling large off-
(yet near-) %change made by MH
equator dips or holes. As shown in Fig. \ref{Figure3CircularSurfaceMill}, reaching the single axial mode regime makes engineered microbubbles effectively respond as high-Q 2D microring resonators directly integrated in a 3D hollow microdevice. However, even in the case of tapered (circular) patterns, the filtering effect comes at the price of a significant drop in the WGMs' Q-factors by more than one order of magnitude. This contradicts the negligible impact of the large hole on the lowest-order axial modes that we expected based on our simulation results (Fig. \ref{Figure1ModeFilterCOMSOLAnalysis}). Despite being well-confined to the outer surface of the resonator, i.e., in the radial direction, the field distributions of microbubble WGMs extend relatively far away from the equator in the axial direction (see, for example, Fig. \ref{Figure1ModeFilterCOMSOLAnalysis}(a)), making microbubble WGMs highly sensitive to surface defects. A possible reason for the mode degradation may lie in the increased surface roughness resulting from large patterning of the sidewalls. %of the large pattern sidewalls 
 One cause could be %caused by 
the relatively high ion-beam current used for milling. Moreover, re-deposition of sputtered material on the surrounding area may contribute to the loss increase.

Despite these challenges, our experimental results demonstrate that local scattering caused by a subwavelength shallow slit pattern, milled at a low ion-beam current, is less detrimental than scattering that occurs when WGMs interact over a long range with a large milled pattern. This effect may be attributed to the local reshaping of the WGMs, at least for fundamental axial modes. As experimentally observed (Fig. \ref{Figure5Localization}), the shallow slit pattern seems to provide better transverse confinement of the WGMs near the large filtering pattern, thereby reducing the interaction between the guided light and the large pattern. By causing only limited radiation losses, the shallow slit pattern thus enables an overall improvement to the WGMs' Q-factors.

Our experimental observations revealed that TE modes, rather than TM modes, tend to experience slightly higher Q-factor drops. This is consistent with TE modes being more sensitive to the microbubble surface roughness \cite{borselli2004rayleigh}. Due to the significant overlap between their intensity distribution and the submicrometer-deep slit pattern, higher-order axial TE modes are expected to suffer higher radiation losses compared to their TM counterparts (Fig. \ref{Figure6PressureTuningwithModeFilters}). However, if their intensity distribution is confined within the slit aperture, the fundamental axial TE modes benefit from the reduced scattering loss provided by the shallow slit pattern (Fig. \ref{Figure4FilterAndConfinement}). In principle, submicrometer-deep patterns with mode-specific designs could also enable a lowering of the scattering losses of higher-order modes, by using similar mechanisms or further photonic operations such as local polarization or mode conversions.

\section{CONCLUSION}\label{sec5}

This work demonstrates how WGM spectral and spatial control can be achieved by integrating FIB-enabled mode filtering and reshaping functionalities on prefabricated microbubble cavities. Although mode confinement and spectral engineering have already been demonstrated in solid and externally controlled WGM systems, deterministic axial-mode filtering in hollow microbubble resonators through localized geometric perturbation has been largely ignored. We have demonstrated that large, off-equator, tapered patterns,  such as circular dips or holes, provide efficient mode filtering, though they also introduce unwanted scattering losses. Additional shallow patterns, such as slits fabricated along the light path, can help reduce those losses by locally confining the WGMs, hence lowering their interactions with large filtering patterns. The benefits of this precision-engineering approach are demonstrated through partial recovery of high Q-factors (close to $10^5$) when using submicrometer-deep slit patterns, especially in the case of fundamental axial TE modes. Taking advantage of the airtight nature of sealed microbubble cavities, we further demonstrated the potential use of such precision-engineered microbubble cavities by achieving mode-mixing-free pressure-tuning of the cavity resonances over a long wavelength range (0.1 FSR). Further engineering of the microbubble cavities could enable even more precise isolation and customization of selected WGMs, leading to the development of hollow cavity devices with more advanced light emission and sensing capabilities.

\section*{APPENDIX: MATERIALS AND METHODS}

\subsection*{Microbubble Fabrication}\label{app1}%
Microbubbles were fabricated  by shining a counter-propagating, focused  C$\text{O}_\text{2}$ laser (Synrad series, 48-2KWM with a maximum power, $P_{max}=25$~W) onto a glass capillary (Polymicro Technologies TSP250350) with inner/outer diameters of 250/350~{\textmu}m \cite{watkins2011single,madugani2016linear}. We used 12\% of $P_{max}$ to remove the outer protective coating from the capillary. The capillary was then  pulled via a stepper motor stage while 20\% $P_{max}$ was shone on it. This initial tapering stage yielded a capillary diameter 10-15$\%$ of its original value. Next, we shone the C$\text{O}_\text{2}$ laser at 30\% $P_{max}$ onto a portion of the tapered capillary while pressurizing it with $\text{N}_\text{2}$ gas (3 bar pressure) to blow it into a bubble shape. By precisely controlling the C$\text{O}_\text{2}$ laser power, the desired microbubble diameter was achieved.

\subsection*{Microbubble FIB Milling}\label{app2}%
For imaging or structuring glass materials using an SEM or FIB milling, respectively, a conductive material coating was used to avoid charge buildup. Here, the microbubbles were sputter-coated, at room temperature, with a 50~nm  layer of Indium Tin Oxide (ITO) on the outer surface. ITO was chosen because its refractive index is very close to that of  fused silica. The microbubbles were then milled with an SEM-FIB (FEI Helios G3 UC). For larger structure milling, such as needed for off-equatorial mode filters, a 0.77~nA current setting was used, whereas for finer submicron pattern writing, a standard 22~pA current setting was used. Initially, we milled a 4~{\textmu}m$^2$, 100~nm deep structure on the bubble surface at $\sim 18$~{~{\textmu}m} away from the equator and used this to adjust the focus of the FIB. We then milled the larger mode filter and submicron structures on the microbubble surface.

\subsection*{Spectral Characterization}\label{app3}%
A fiber taper was used to couple light into the microbubble resonators and excite the optical modes. A 125~{~{\textmu}m} diameter single-mode fiber (Thorlabs, 1550BHP), was tapered down to a 1.5~{~{\textmu}m diameter using a hydrogen-oxygen flame-brushing fiber pulling rig \cite{ward2014contributed}. During measurements the fiber taper was in contact coupling with the microbubble resonators.  A photodetector (PDB450C) connected to a digital storage oscilloscope (Moku:Lab) was used to record the fiber transmission while scanning the frequency of a  1550~nm tunable laser (New Focus Velocity TLB-6728). For the FIB-processed microbubbles, better mode-matching was achieved via the fiber taper over an angular range of about $52^{ \circ }$, defined in relation to  axial rotation of the microbubble from the slot region. Each IR image was obtained either by a tunable laser fine-scanning over a narrow tuning range across the transmission line or by the laser being proportional–integral–derivative (PID)-locked to it.  A time-averaged far-field IR camera was used to capture optical field distributions to identify the modes. Due to the orientation limitation, the images using an IR camera were obtained by focusing on the slit area, at the top edge, and slightly below the top edge of the milled resonator.

\section*{Author Contributions}

All authors contributed to the article writing. R.M. conceived the idea, contributed to project planning, device fabrication, experimental investigations and analysis. A.J. equally contributed to the device fabrication, experimental investigations and analysis. C.P. contributed to the simulations, theoretical investigations and analysis. M.O. contributed to the device fabrication. M.H. contributed to the theoretical investigations,
simulation advice and discussions. S.N. contributed to project planning, execution, article writing, experimental advice, discussions, and overview of the project.

\section*{Acknowledgment}
The authors would like to thank  the Engineering, Scientific Imaging, and Scientific Computing \& Data Analysis Sections of the Okinawa Institute of Science and Technology Graduate University (OIST) and K. Karlsson for technical assistance.

\section*{Funding}
This work was partly funded by the Okinawa Institute of Science and Technology Graduate University (OIST), by JSPS KAKENHI Grant No. JP24K08290 and by the Japanese Cross-Ministerial Strategic Innovation Promotion Program (SIP) under Grant No. JPJ012367).

\section*{Conflicts of Interest}
The authors declare no conflicts of interest.

\section*{Data Availability}
The data are available from the authors upon reasonable request.

\bibliography{References}

\end{document}